\documentclass[runningheads]{llncs}
\usepackage[T1]{fontenc}
\usepackage{graphicx}
\usepackage{booktabs}
\usepackage{amsmath}
\usepackage{mdframed}
\usepackage[
  square,        
  comma,         
  numbers,       
  sort&compress, 
]{natbib}

\usepackage{subcaption}

\begin{document}
\title{ACSE-Eval: Can LLMs threat model real-world cloud infrastructure?}

%
%

\author{Sarthak Munshi \and
Swapnil Pathak \and
Sonam Ghatode
\and
Thenuga Priyadarshini
\and
Dhivya Chandramouleeswaran
\and
Ashutosh Rana
}

\authorrunning{S. Munshi et al.}
%
\maketitle              
\begin{abstract}
While Large Language Models (LLMs) have shown promise in cybersecurity applications, their effectiveness in identifying security threats within cloud deployments remains unexplored. This paper introduces AWS Cloud Security Engineering (ACSE)-Eval, a novel dataset for evaluating LLMs’ cloud security threat modeling capabilities. ACSE-Eval contains 100 production-grade AWS deployment scenarios, each featuring detailed architectural specifications, Infrastructure as Code (IaC) implementations, documented security vulnerabilities, and associated threat modeling parameters. Our dataset enables systemic assessment of LLMs’ abilities to identify security risks, analyze attack vectors, and propose mitigation strategies in cloud environments. Our evaluations on ACSE-Eval demonstrate that GPT-4.1 and Gemini 2.5 Pro excel at threat identification, with Gemini 2.5 Pro performing optimally in 0-shot scenarios and GPT-4.1 showing superior results in few-shot settings. While GPT-4.1 maintains a slight overall performance advantage, Claude 3.7 Sonnet generates the most semantically sophisticated threat models but struggles with threat categorization and generalization. To promote reproducibility and advance research in automated cybersecurity threat analysis, we open-source our dataset\footnote{https://huggingface.co/datasets/ACSE-Eval/ACSE-Eval}, evaluation metrics, and methodologies.

\end{abstract}
\section{Introduction}
Large Language Models (LLMs) have demonstrated promising performance in cybersecurity tasks such as vulnerability detection and code analysis \cite{chen2023diversevulnewvulnerablesource} \cite{ferrag2025securefalconautomatedsoftwarevulnerability}. However, their ability to perform architectural threat assessments in complex cloud environments remains under explored. While effective at identifying source-level issues \cite{sourcellm}, LLMs have yet to prove they can reason about service interactions, trust boundaries, and multi-resource configurations typical of modern cloud systems. The urgency of this research is underscored by the evolving threat landscape in cloud security. In 2024, cloud breaches have reached alarming levels, with 79\% of cloud-based enterprises reporting at least one incident, and about 25\% of organizations expressed uncertainty about undetected threats \cite{css25,sentinelone24}. More concerning is that 82\% of these breaches originated from architectural design flaws or misconfigurations, highlighting a critical gap in current security approaches. 

Traditional approaches to threat modeling, exemplified by frameworks like MITRE ATT\&CK \cite{mitre252} and STRIDE \cite{stride-framework}, have been enhanced with cloud-specific considerations. However, these frameworks, while comprehensive in mapping adversarial tactics and techniques, have yet to fully integrate with LLM capabilities for automated threat assessment. The complexity of modern cloud architectures, combined with the dynamic nature of Identity and Access Management (IAM) and Zero Trust requirements, demands more sophisticated threat modeling capabilities that can adapt to rapidly evolving security challenges.

\begin{figure}[htbp]
    \centering
    
    \begin{subfigure}{\textwidth}
        \centering
        \includegraphics[width=\linewidth, keepaspectratio]{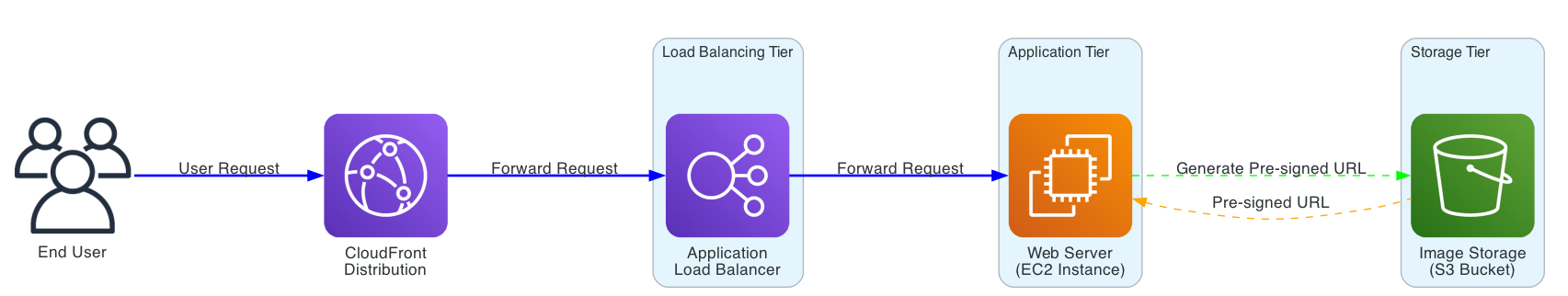}
        \label{fig:top}
    \end{subfigure}
    
    \vspace{0.5em} 
    
    \begin{minipage}{0.48\textwidth}
        \begin{mdframed}[linewidth=0.5pt]
            \tiny
            \begin{verbatim}
from diagrams.aws.network import ELB, CloudFront
from diagrams.aws.general import Users
from diagrams.aws.compute import EC2
from diagrams.aws.storage import S3
from diagrams import Diagram, Cluster, Edge

...

with Diagram("AWS Website Architecture", ...) as diag:
    user = Users("End User")
    cdn = CloudFront("CloudFront Distribution")
    
    with Cluster("Load Balancing Tier"):
        alb = ELB("Application Load Balancer")
    
    with Cluster("Application Tier"):
        web_server = EC2("Web Server (EC2 Instance)")
    
    with Cluster("Storage Tier"):
        image_bucket = S3("Image Storage (S3 Bucket)")
    
    # Define main flows
    user >> Edge(color="blue", label="User Request") \
    >> cdn >> alb >> web_server
    web_server >> Edge(color="green", \
    label="Pre-signed URL ops") >> image_bucket
    ...
            \end{verbatim}
        \end{mdframed}
    \end{minipage}%
    \hspace{0.04\textwidth}%
    \begin{minipage}{0.48\textwidth}
        \begin{mdframed}[linewidth=0.5pt]
            \tiny
            \begin{verbatim}
import * as cdk from 'aws-cdk-lib';
import { ec2, ... } from 'aws-cdk-lib/aws-*';

export class SimpleImageWebAppStack extends cdk.Stack {
  constructor(...) {
    super(scope, id, props);

    // VPC with public/private subnets
    const vpc = new ec2.Vpc(this, 'AppVpc', {
      maxAzs: 2,
      subnetConfiguration: [,
        ...
      ]
    });

    // Core infrastructure components
    const imageBucket = new s3.Bucket(...);

    const webServer = new ec2.Instance(...);

    const alb = new elbv2.ApplicationLoadBalancer(...);

    const dist = new cloudfront.Distribution(...);
  }
}
...
            \end{verbatim}
        \end{mdframed}
    \end{minipage}
    
    \caption{Architecture and implementation of a sample cloud infrastructure scenario (a simple S3 pre-signed URL service for image handling). \emph{Top:} System architecture diagram. \emph{Bottom-left:} Diagram-as-Code in Python. \emph{Bottom-right:} AWS CDK IaC in TypeScript.}
    \label{fig:full}
\end{figure}

Recent initiatives in the LLM-enabled threat-modeling space are steps in the right direction. Elsharef et al. \cite{ndsssymposium} explores the practical applicability of an LLM assisted threat model but lacks a comprehensive evaluation procedure. Projects such as Auspex \cite{auspex} and ThreatModeling-LLM \cite{yang2024} aim to address this. However, they are focused on industry-specific infrastructure such as banking. Existing datasets primarily evaluate LLMs using Capture-the-Flag (CTF) challenges and exploitation tasks \cite{cyberseceval2,penheal,cyberseceval3,occult}. While valuable, these gamified, bounded tasks fail to reflect the complexity of real-world cloud infrastructure. Dynamic testing frameworks \cite{phuong24,ctibench} and red teaming initiatives like DARPA AIxCC broaden this scope but still emphasize exploit generation over architectural reasoning. Moreover, many tasks risk conflating memorization with genuine security understanding. Recently, there have been efforts to benchmark general cybersecurity knowledge \cite{cybermetric,seceval,secqa}, threat intelligence \cite{ctibench}, and IaC security \cite{kon2024iaceval}, but they do not address holistic threat modeling across cloud architectures and their corresponding IaC artifacts. 

To address these limitations, we present \textbf{ACSE-Eval}, a dataset designed to evaluate LLMs' threat modeling capabilities. Our contributions include: \emph{a) }a curated dataset of 100 real-world AWS architecture diagrams with Diagrams-as-Code \cite{diagrams} and Infrastructure-as-Code implementations using AWS CDK \cite{aws-cdk}, \emph{b)} expert-generated threat models aligned with STRIDE \cite{shostack2014threat}, ATT\&CK \cite{mitre252}, and OWASP Top 10 \cite{owasp252}, \emph{c)} a multi-dimensional evaluation framework assessing threat identification, vulnerability analysis, and mitigation suggestions, and \emph{d)} an open-source release of the dataset and evaluation toolkit to promote further research. Spanning 100 human-annotated threat scenarios and over 2,500 hours (about 3 months) of expert effort, our dataset targets AWS, representing 31\% of the cloud market \cite{techcrunch} and includes use cases ranging from simple web apps to multi-region, and hybrid deployments. 

In addition to advancing the field of cybersecurity evaluations for LLMs via the open-source dataset, our work aims to address the following key research questions:

\emph{R1: Can current language models effectively identify infrastructure security issues via IaC?}\\
The question aims to evaluate both the accuracy and reliability of LLMs in identifying potential security misconfigurations, compliance violations, and architectural weaknesses in infrastructure definitions expressed through code (AWS CDK). We measure the coverage of vulnerability detection, and the comprehensiveness of the security analysis. This investigation is particularly relevant given the increasing adoption of IaC in cloud deployments and the potential for automated security analysis to enhance infrastructure security at scale while reducing human error in security reviews.
\\\\
\emph{R2: Can the threat modeling capabilities of language models be enhanced through the integration of visual-esque aids or relationship-defining tools, specifically using Diagrams-as-Code?}\\
We explore the potential for improving the threat modeling capabilities of LLMs by incorporating codified visual representation techniques, particularly focusing on Diagrams-as-Code or Component Relationship Context (CRC). The inquiry seeks to determine whether the addition of tools that define relationships between system components can enhance an LLM's ability to perform threat modeling. CRC, which allows for the programmatic creation and manipulation of visual diagrams, could provide LLMs with a more structured and explicit representation of system architectures, data flows, and component interactions. This approach might enable LLMs to better understand complex systems, identify potential attack vectors, and reason about security implications more effectively than when working with text descriptions or IaC alone. The question aims to assess whether this integration could lead to more thorough threat identification.
\\\\
\emph{R3: How well can language models provide contextually appropriate security recommendations?}\\
We also examine LLMs' ability to generate security recommendations that are appropriately tailored to specific contexts, environments, and constraints. The question highlights whether LLMs can effectively consider factors such as the application domain, technical limitations, and resource constraints when proposing security controls or mitigations. The term \emph{contextually appropriate} is crucial here, as it goes beyond merely identifying security issues to assess whether the suggested solutions are practical, implementable, and aligned with the specific needs and circumstances of the target environment.

\section{Methodology and dataset}

ACSE-Eval introduces a structured methodology for evaluating the capability of LLMs to perform threat modeling on real-world cloud architectures. The work spans three stages: (1) data generation, (2) expert analysis, (3) evaluation metrics, and LLM performance assessment. This is explained in the following sections. 
\subsection{Dataset Generation Workflow}
\begin{figure}[t]
    \centering
    \includegraphics[width=0.95\linewidth, keepaspectratio]{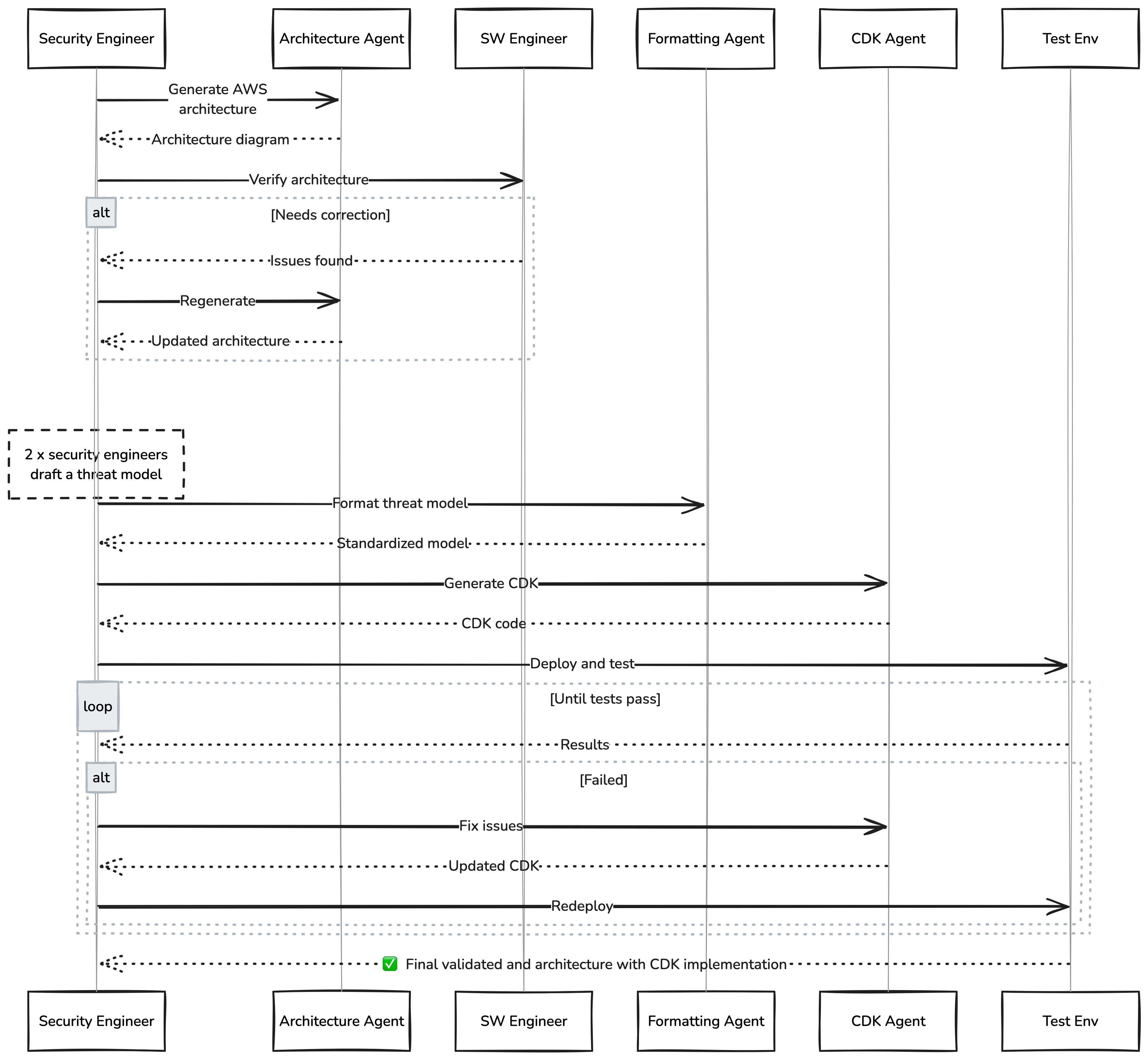}
    \caption{Methodology used for generating the ACSE-Eval dataset.}
    \label{fig:methodology}
\end{figure}

The process (illustrated by Figure \ref{fig:methodology}) begins with the initial architecture generation phase, where a Security Engineer interfaces with a specialized \emph{Architecture Agent} to create an AWS architecture. This interaction is designed to incorporate security requirements and best practices from the outset. The generated architecture then undergoes scrutiny from a Software Engineer, who evaluates it for technical feasibility, compliance with organizational standards, and potential implementation challenges. If any issues are identified during this review, a feedback loop is initiated where the Security Engineer works with the \emph{Architecture Agent} to regenerate the architecture with specific corrections. This cycle of generation, review, and refinement continues until the architecture meets all security requirements and receives approval.

Following architectural approval, the methodology moves into a crucial collaborative threat modeling phase. This stage brings together two Security Engineers to perform comprehensive threat analysis, leveraging their combined expertise and diverse perspectives to identify potential security risks and vulnerabilities. The resulting threat model is then processed by a specialized \emph{Formatting Agent}, which standardizes it into a consistent JSON file. The next phase involves translating the approved architecture into actual infrastructure code. A \emph{CDK Agent} takes on this task, generating AWS CDK code that implements the architecture while maintaining the security controls and configurations specified in the design. This automation helps reduce human error in the implementation phase while ensuring consistency between the architectural design and the actual infrastructure code.

\begin{figure}
    \centering
    \includegraphics[width=0.6\linewidth, keepaspectratio]{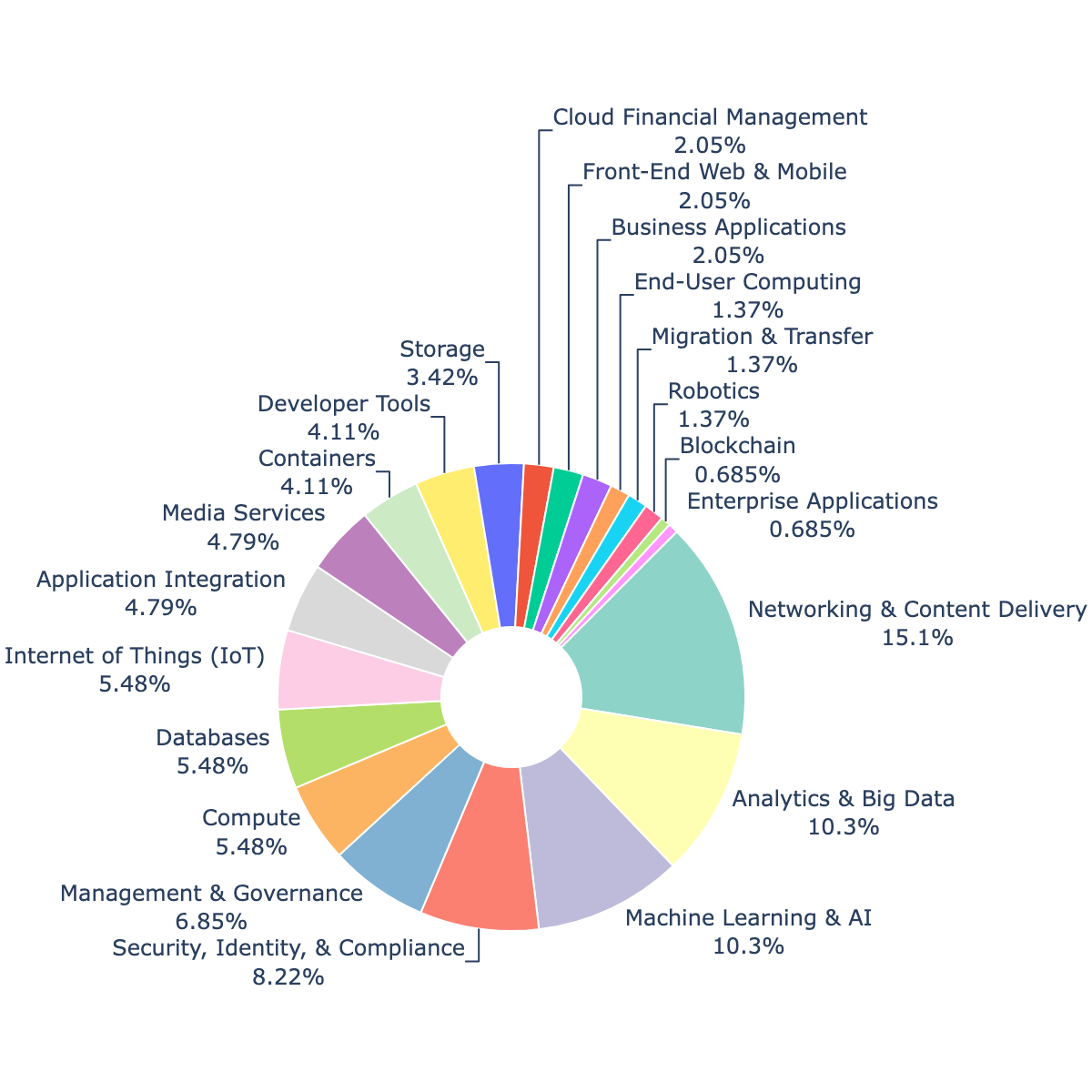}
    \caption{Category-wise breakdown of the AWS services that are part of ACSE-Eval.}
    \label{fig:aws_services}
\end{figure}

The methodology then enters a testing and validation loop. The generated CDK code is deployed to a dedicated test environment. This testing phase is designed to ensure the infrastructure functions as intended. When tests fail, a structured manual remediation process begins. The Security Engineer works with the \emph{CDK Agent} to address specific issues, generating updated CDK code that is then redeployed to the test environment. This cycle of testing, feedback, and improvement continues until all tests pass successfully.

\subsection{Dataset Overview}

The dataset reflects real-world diversity in deployment patterns, application domains, and security postures. Architectures vary in complexity (from minimal viable products to highly distributed systems), incorporate a wide range of AWS services, and include both secure and intentionally misconfigured configurations.

\vspace{0.5em}
\textbf{Architecture Distribution.} The dataset spans 12 different infrastructure categories: Data Processing \& Analytics (15.05\%), AI/ML \& Compute Platforms (11.83\%), Business Applications (11.83\%), Infrastructure \& Networking (8.60\%), Specialized Systems (8.60\%), Serverless Architectures (7.53\%), Media \& Content Services (7.53\%), Security \& Identity (7.53\%), Multi-Region \& High Availability (6.45\%), IoT \& Connected Systems (5.38\%), Gaming (5.38\%), and Collaboration \& Communication (4.30\%).

\begin{figure}
    \centering
    \includegraphics[width=\linewidth, keepaspectratio]{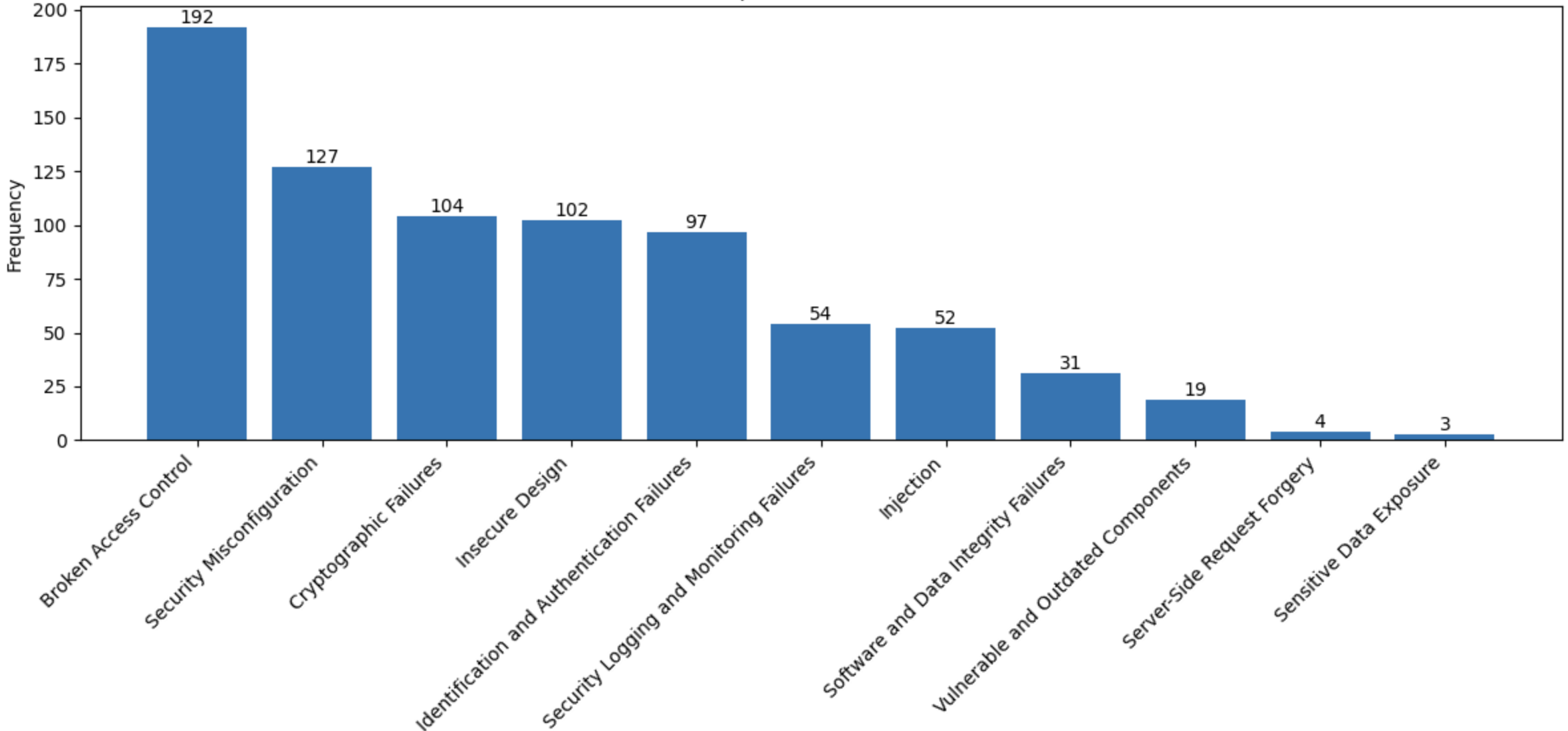}
    \caption{Distribution of OWASP Top 10 threats in the ACSE-Eval dataset.}
    \label{fig:owasp_distribution}
\end{figure}

\textbf{Service Coverage.} The benchmark covers 146 distinct AWS services across compute, storage, networking, identity, analytics, ML, IoT, and security among other domains. The distribution is illustrated by Figure \ref{fig:aws_services}. This breadth challenges LLMs to reason across diverse primitives, policies, and interactions.

\textbf{Threat Models.} Threat models within ACSE-Eval span 115 distinct threats, spread across the STRIDE, ATT\&CK, and OWASP Top 10 frameworks. Threats were derived from architecture analysis, IaC inspection, and attack vector mapping. Scenarios include realistic flaws, such as misconfigurations (e.g., open S3 buckets), missing controls (e.g., absent logging), design flaws (e.g., flat trust boundaries), implementation bugs in CDK, and compliance violations (e.g., no encryption at rest). This diversity of flaws ensures LLMs must reason about vulnerabilities at the architectural layers.

\begin{figure}[t]
    \centering
    \begin{subfigure}[b]{0.49\linewidth}
        \centering
        \includegraphics[width=\textwidth, keepaspectratio]{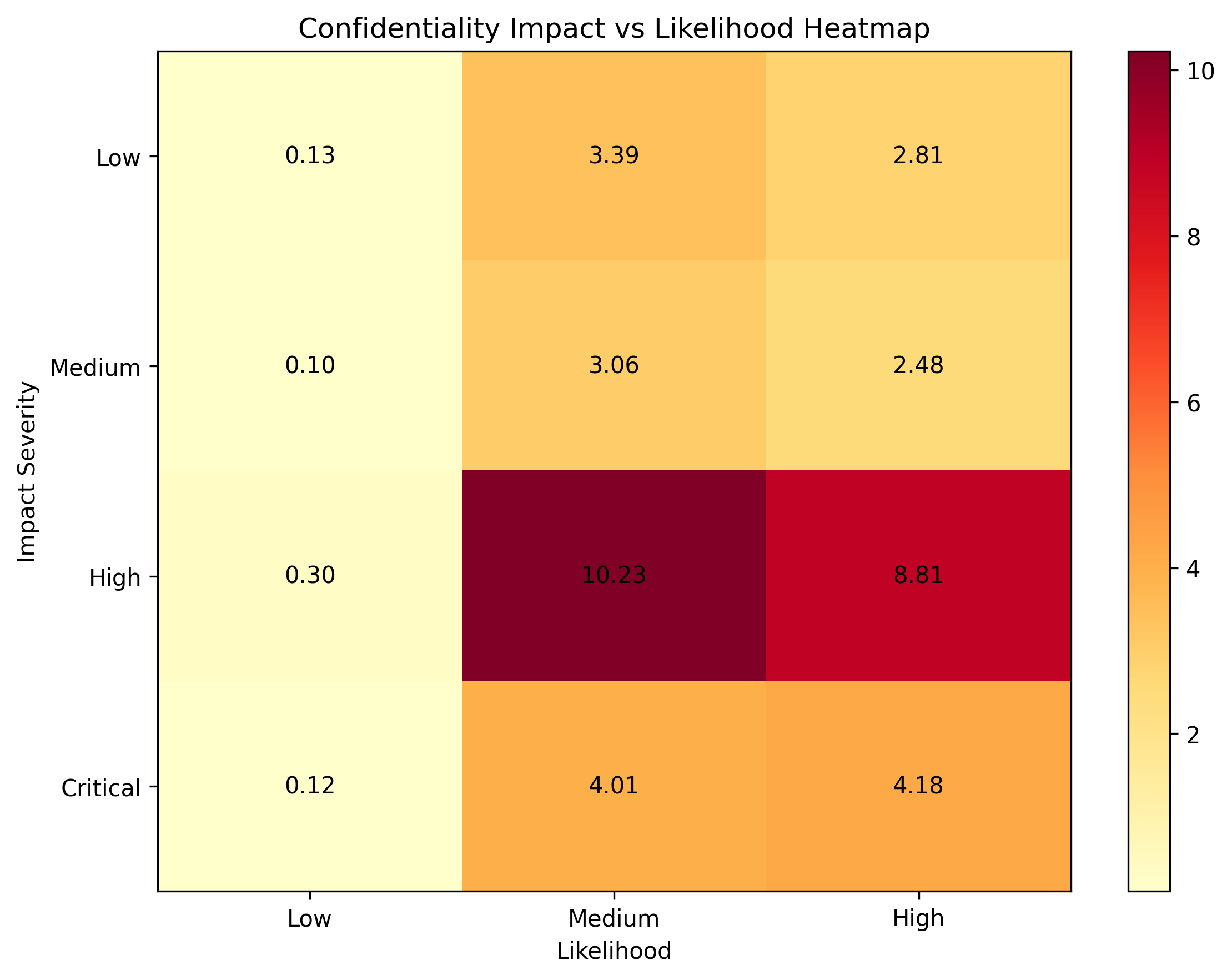}
        \label{fig:image1}
    \end{subfigure}%
    \begin{subfigure}[b]{0.49\linewidth}
        \centering
        \includegraphics[width=\textwidth, keepaspectratio]{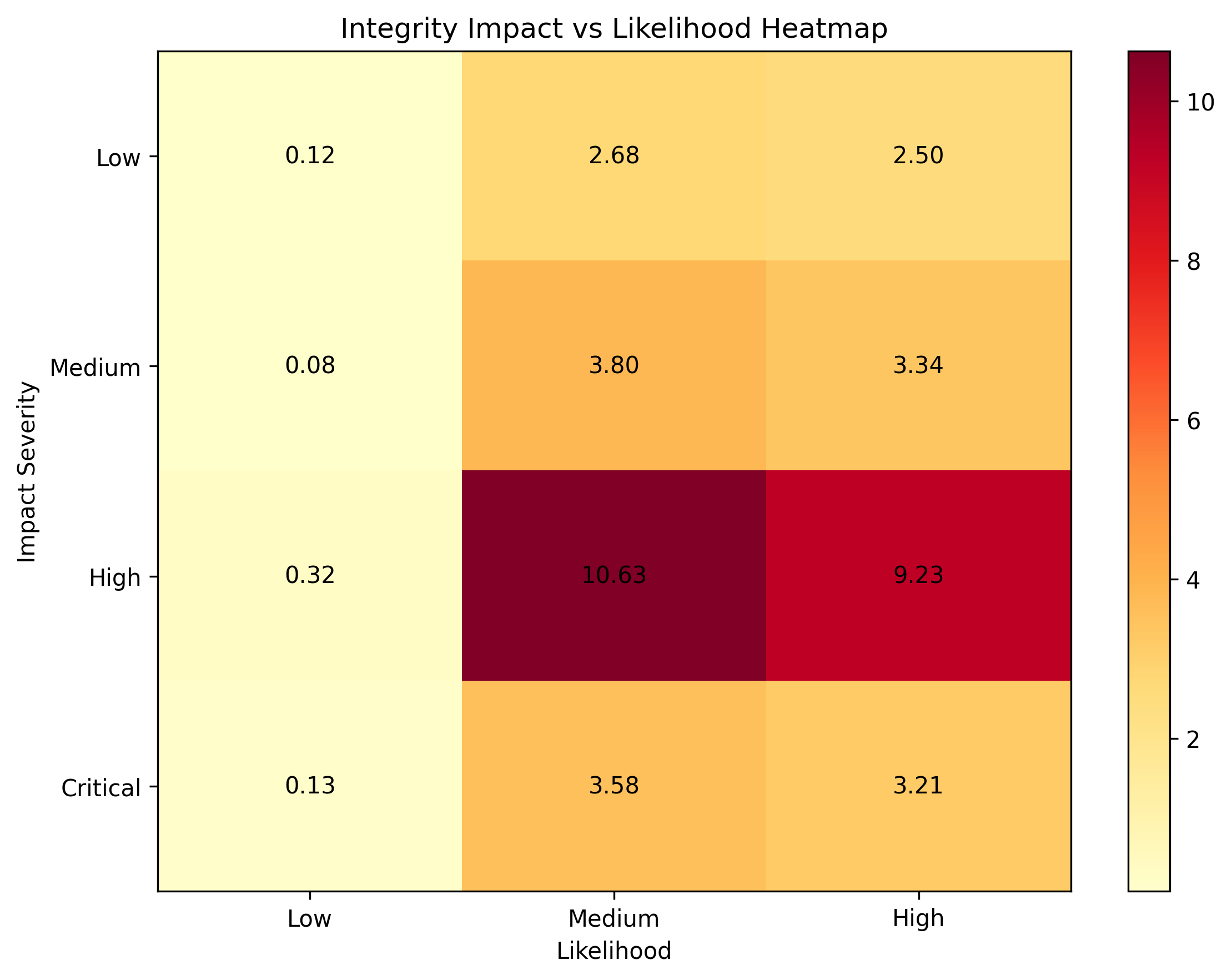}
        \label{fig:image2}
    \end{subfigure}
    
    \vspace{0.5cm}
    
    \begin{subfigure}[b]{0.49\linewidth}
        \centering
        \includegraphics[width=\textwidth, keepaspectratio]{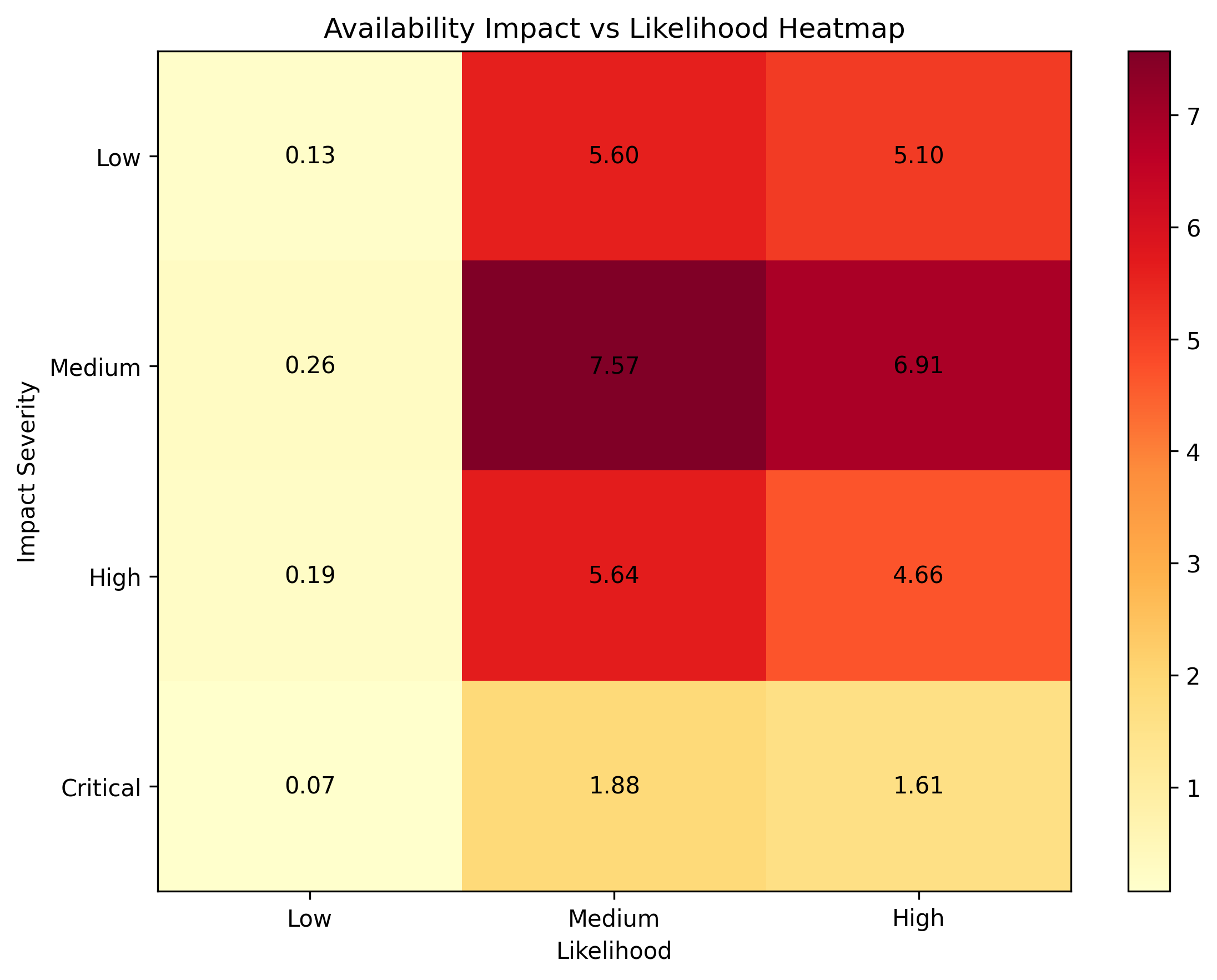}
        \label{fig:image3}
    \end{subfigure}
    \caption{Heatmaps showing the relationship between threat likelihood and CIA (Confidentiality, Integrity, Availability) impact across all architectures evaluated in ACSE-Eval.}
    \label{fig:category_services}
\end{figure}

\textbf{Component Relationship Context (CRC).} Each threat model and IaC implementation is accompanied by a PNG architecture diagram and its corresponding Python-based diagram generation code \cite{diagrams}. While our current scope focuses on textual inputs and evaluation, these architectural visualizations serve as valuable artifacts for establishing relationships between various system components. We hypothesize that providing LLMs with this component relationship context, alongside the deployment code, could enhance the quality of generated threat models. Future research could extend this work by incorporating visual (multi-modal) in-context learning techniques, potentially leading to more comprehensive and accurate threat modeling capabilities.

\section{Experiments}

\subsection{Implementation and Tooling}

Our evaluations leverage the managed LLM APIs provided by Amazon 
 \cite{aws2023bedrock}, Anthropic \cite{anthropic2024claude35sonnet}, Google \cite{google2023gemini}, and OpenAI \cite{openai2023gpt4}. The comprehensive experiment incurred a cost of approximately \$500. The evaluation framework is implemented using Inspect AI, an open-source platform for LLM assessments developed by \cite{InspectAI2024}. During our analysis, we employed specialized packages for text similarity measurements, including \texttt{rouge} for calculating Rouge-L scores \cite{lin-2004-rouge} and \texttt{sentence-transformers} for computing semantic (cosine) similarity \cite{reimers-2019-sentence-bert}. To ensure transparency and reproducibility, the source code for both the experiment and evaluation procedures is freely available at \url{https://github.com/ACSE-Eval/acse-eval-experiments} under the MIT License.

Additionally, our dataset generation agents (\emph{Architecture}, \emph{CDK}, and \emph{Formatting}) employ a hybrid approach, combining human intervention with the capabilities of Claude 3.5 Sonnet. For evaluations, we analyze Claude 3.5 Haiku ({\itshape claude-3-5-haiku-20241022}), Claude 3.7 Sonnet (\emph{claude-3-7-sonnet-20250219}), GPT 4.1 (\emph{gpt-4.1-2025-04-1}), GPT 4o (\emph{gpt-4o-2024-08-06}), Deepseek V3, and Gemini 2.5 Pro (\emph{gemini-2.5-pro-preview-05-06}). We exclude Claude 3.5 Sonnet (\emph{claude-3-5-sonnet-20240620, claude-3-5-sonnet-20241022}) from our evaluations to prevent implicit bias due to its use with the agents. We set the temperature of LLMs at 0 and top\_p = 1 to obtain more deterministic responses. Each task is evaluated 3-times on a zero-shot and 3-shot
prompt template with instruction of LLMs to act as a cloud security engineering expert. For each 3-shot task run, we seed and extract the 3-shot examples at random. We calculate the mean and standard error mean for every metric in the following sections. 

\subsection{Evaluating LLMs performance on ACSE-Eval}

\vspace{0.5em}
\subsubsection{Framework Coverage} 
We assess the accuracy of threat identification by calculating the proportion of reference threats successfully captured in the model-generated threat model. To evaluate this, we introduce a metric called Threat Framework Coverage (TFC), which measures the recall percentage ($\frac{TP \times 100}{TP + FN}$) for each security framework. This calculation provides a quantitative measure of how effectively the model identifies and categorizes threats within established security frameworks.

\begin{table}[htb]
\centering
\caption{TFC evaluation for STRIDE.}
\begin{tabular}{lcccccc}
\toprule
& \multicolumn{2}{c}{\textbf{CDK}} & \multicolumn{2}{c}{\textbf{IaC + CRC}} \\
\cmidrule(lr){2-3} \cmidrule(lr){4-5}
\textbf{Model} & \textbf{0-shot} & \textbf{3-shot} & \textbf{0-shot} & \textbf{3-shot} \\
\midrule
Claude 3.5 Haiku & $56.8 \pm 2.9$ & $32.3 \pm 4.37$ & $57.6 \pm 2.96$ & $19.6 \pm 3.59$  \\
Claude 3.7 Sonnet & $80.2 \pm 1.52$ & $87.8 \pm 1.16$ & $84.7 \pm 1.3$ & $88.8 \pm 1.14$ \\
GPT 4o & $60.9 \pm 1.86$ & $78.8 \pm 1.66$ & $66.3 \pm 2.03$ & $84.2 \pm 1.45$ \\
GPT 4.1 & $95.1 \pm 0.756$ & $\mathbf{96.8 \pm 0.709}$ & $97.4 \pm 0.659$ & $\mathbf{98.7 \pm 0.483}$ \\
DeepSeek Chat V3 & $94.4 \pm 0.908$ & $93.7 \pm 1.02$ & $94.8 \pm 0.914$ & $95.7 \pm 0.791$ \\
Gemini 2.5 Pro & $\mathbf{96.2 \pm 0.835}$ & $92.6 \pm 1.09$ & $\mathbf{98.4 \pm 0.527}$ & $92.9 \pm 1.12$ \\
\bottomrule
\end{tabular}
\end{table}

\begin{table}[htb]
\centering
\caption{TFC evaluation for OWASP Top 10.}
\begin{tabular}{lcccccc}
\toprule
& \multicolumn{2}{c}{\textbf{IaC}} & \multicolumn{2}{c}{\textbf{IaC + CRC}} \\
\cmidrule(lr){2-3} \cmidrule(lr){4-5}
\textbf{Model} & \textbf{0-shot} & \textbf{3-shot} & \textbf{0-shot} & \textbf{3-shot} \\
\midrule
Claude 3.5 Haiku & $39.0 \pm 2.4$ & $26.7 \pm 3.68$ & $40.5 \pm 2.54$ & $15.4 \pm 3.0$\\
Claude 3.7 Sonnet & $70.4 \pm 1.81$ & $74.6 \pm 1.62$ & $70.2 \pm 1.74$ & $71.8 \pm 1.78$\\
GPT 4o & $42.0 \pm 1.86$ & $56.3 \pm 1.78$ & $41.6 \pm 2.0$ & $62.4 \pm 1.75$\\
GPT 4.1 & $81.9 \pm 1.73$ & $\mathbf{85.9 \pm 1.73}$ & $88.1 \pm 1.53$ & $\mathbf{90.0 \pm 1.47}$\\
DeepSeek Chat V3 & $70.4 \pm 1.62$ & $72.8 \pm 1.56$ & $68.4 \pm 1.72$ & $74.8 \pm 1.58$\\
Gemini 2.5 Pro & $\mathbf{85.9 \pm 1.38}$ & $83.5 \pm 1.64$ & $\mathbf{89.1 \pm 1.33}$ & $85.1 \pm 1.49$ \\
\bottomrule
\end{tabular}
\end{table}

\begin{table}[htb]
\centering
\caption{TFC evaluation for MITRE ATT\&CK.}
\begin{tabular}{lcccccc}
\toprule
& \multicolumn{2}{c}{\textbf{IaC}} & \multicolumn{2}{c}{\textbf{IaC + CRC}} \\
\cmidrule(lr){2-3} \cmidrule(lr){4-5}
\textbf{Model} & \textbf{0-shot} & \textbf{3-shot} & \textbf{0-shot} & \textbf{3-shot} \\
\midrule
Claude 3.5 Haiku & $17.7 \pm 1.68$ & $14.2 \pm 2.39$ & $19.3 \pm 1.6$ & $8.24 \pm 1.68$\\
Claude 3.7 Sonnet & $36.2 \pm 1.75$ & $40.6 \pm 1.92$ & $38.9 \pm 1.92$ & $42.3 \pm 1.95$\\
GPT 4o & $17.4 \pm 1.31$ & $27.4 \pm 1.9$ & $19.1 \pm 1.36$ & $27.0 \pm 1.67$\\
GPT 4.1 & $44.3 \pm 1.75$ & $\mathbf{49.5 \pm 2.12}$ & $42.8 \pm 1.93$ & $\mathbf{49.9 \pm 2.07}$\\
DeepSeek Chat V3 & $36.1 \pm 1.7$ & $42.5 \pm 1.8$ & $38.5 \pm 1.81$ & $41.0 \pm 1.86$\\
Gemini 2.5 Pro & $\mathbf{45.9 \pm 1.78}$ & $46.1 \pm 2.03$ & $\mathbf{46.9 \pm 1.89}$ & $47.4 \pm 1.85$ \\
\bottomrule
\end{tabular}
\end{table}

Analysis of the results reveal distinct performance patterns across models. GPT 4.1 demonstrates superior performance with 3-shot generation, while Gemini 2.5 Pro excels in 0-shot scenarios. Notably, 3-shot generation proves counterproductive for Gemini 2.5 Pro, occasionally resulting in decreased Threat Framework Coverage. The smaller models, including Claude 3.5 Haiku and GPT4o, show particular difficulty in accurately classifying MITRE ATT\&CK codes. When evaluating baseline performance without considering Component Relationship Context (CRC) or the number of examples, Gemini 2.5 Pro demonstrates consistent superiority across all three security frameworks, establishing itself as the most effective model for threat identification and classification. On the other hand, including either CRC or variations in shot counts in our analysis, GPT 4.1 demonstrates better performance.

\vspace{0.5em}
\subsubsection{Text Similarity} To account for linguistic variation in threat descriptions, we compute and evaluate certain text similarity metrics. These allow us to credit partial correctness and stylistic variation, crucial in an open-ended reasoning task, especially threat mitigation recommendations.

\textbf{ROUGE-L.} This metric measures longest common subsequence between the predicted and reference text.\\
    Let $X$ be the target text and $Y$ be the model output. The ROUGE-L score is calculated as:

\begin{equation}
    \text{ROUGE-L} = \frac{(1 + \beta^2)R_lP_l}{R_l + \beta^2P_l}
\end{equation}

where $R_l = \frac{\text{LCS}(X,Y)}{\text{length}(X)}$ is the LCS-based recall, $P_l = \frac{\text{LCS}(X,Y)}{\text{length}(Y)}$ is the LCS-based precision, $\beta = 1.2$ to favor recall over precision, and $\text{LCS}(X,Y)$ is the length of the Longest Common Subsequence between $X$ and $Y$.

This metric evaluates how well the model output ($Y$) matches the target text ($X$) by finding the longest sequence of words that appears in both texts in the same order.

\textbf{Semantic Similarity.} This metric measures the cosine of the angle between the embedding vectors, providing a value between -1 and 1, where 1 indicates perfect similarity, 0 indicates no similarity, and -1 indicates perfect dissimilarity. It captures the semantic closeness of the model output to the target text, regardless of exact word matching. In order to calculate this, we transform the threat model JSON content (for both the model output and target text) to vectors and compute their cosine similarity.

Let $X$ be the target text and $Y$ be the model output. The Semantic Similarity score is calculated as:

\begin{equation}
    \text{Semantic Similarity} = \cos(\theta) = \frac{E(X) \cdot E(Y)}{\|E(X)\| \|E(Y)\|}
\end{equation}

where $E(X)$ and $E(Y)$ are the sentence-transformer embeddings of $X$ and $Y$ respectively, and $\|\cdot\|$ represents the L2 norm (Euclidean length) of the vector.

\begin{table}[htb]
\centering
\caption{ROUGE-L scores for ACSE-Eval.}
\begin{tabular}{lcccccc}
\toprule
& \multicolumn{2}{c}{\textbf{IaC}} & \multicolumn{2}{c}{\textbf{IaC + CRC}} \\
\cmidrule(lr){2-3} \cmidrule(lr){4-5}
\textbf{Model} & \textbf{0-shot} & \textbf{3-shot} & \textbf{0-shot} & \textbf{3-shot} \\
\midrule
Claude 3.5 Haiku & $38.4 \pm 3.47$ & $24.8 \pm 3.81$ & $38.5 \pm 3.27$ & $16.2 \pm 3.34$  \\
Claude 3.7 Sonnet & $70.3 \pm 2.53$ & $\mathbf{79.4 \pm 1.98}$ & $69.6 \pm 2.45$ & $\mathbf{83.7 \pm 1.75}$ \\
GPT 4o & $47.4 \pm 3.33$ & $60.3 \pm 3.12$ & $48.7 \pm 3.27$ & $58.3 \pm 3.23$ \\
GPT 4.1 & $\mathbf{74.4 \pm 2.35}$ & $77.0 \pm 2.35$ & $\mathbf{76.0 \pm 2.31}$ & $78.7 \pm 2.32$ \\
DeepSeek Chat V3 & $61.2 \pm 3.09$ & $74.7 \pm 2.98$ & $61.2 \pm 3.19$ & $79.8 \pm 2.66$ \\
Gemini 2.5 Pro & $71.6 \pm 2.37$ & $74.3 \pm 2.21$ & $74.6 \pm 2.09$ & $74.6 \pm 2.27$ \\
\bottomrule
\end{tabular}
\end{table}

\begin{table}[htb]
\centering
\caption{Semantic Similarity scores for ACSE-Eval.}
\begin{tabular}{lcccccc}
\toprule
& \multicolumn{2}{c}{\textbf{IaC}} & \multicolumn{2}{c}{\textbf{IaC + CRC}} \\
\cmidrule(lr){2-3} \cmidrule(lr){4-5}
\textbf{Model} & \textbf{0-shot} & \textbf{3-shot} & \textbf{0-shot} & \textbf{3-shot} \\
\midrule
Claude 3.5 Haiku & $0.634 \pm 0.006$ & $0.589 \pm 0.008$ & $0.632 \pm 0.006$ & $0.585 \pm 0.010$  \\
Claude 3.7 Sonnet & $\mathbf{0.798 \pm 0.009}$ & $\mathbf{0.825 \pm 0.008}$ & $\mathbf{0.822 \pm 0.007}$ & $\mathbf{0.819 \pm 0.01}$ \\
GPT 4o & $0.642 \pm 0.01$ & $0.792 \pm 0.007$ & $0.648 \pm 0.01$ & $0.805 \pm 0.007$ \\
GPT 4.1 & $0.745 \pm 0.008$ & $0.799 \pm 0.008$ & $0.747 \pm 0.009$ & $\mathbf{0.823 \pm 0.008}$ \\
DeepSeek Chat V3 & $0.739 \pm 0.009$ & $0.74 \pm 0.008$ & $0.748 \pm 0.009$ & $0.736 \pm 0.007$ \\
Gemini 2.5 Pro & $0.742 \pm 0.008$ & $0.756 \pm 0.006$ & $0.752 \pm 0.007$ & $0.756 \pm 0.007$ \\
\bottomrule
\end{tabular}
\end{table}

We observe that, while Claude 3.7 Sonnet consistently achieves the highest Semantic Similarity scores, GPT 4.1 demonstrates superior performance in ROUGE-L scoring specifically under 0-shot conditions. A higher Semantic Similarity score coupled with a relatively lower ROUGE-L score (as in the case of Claude 3.7 Sonnet) indicates that while the model captures the core meaning and concepts of the reference text, it expresses these ideas using different vocabulary and sentence structures. In threat modeling contexts, this could mean the model identifies threats correctly but describes them using alternative, though semantically equivalent language or format.

\vspace{0.5em}
\subsubsection{Comprehensiveness} To evaluate the models' precision in technical classifications and affected threat surface components, we assess their ability to correctly identify Common Weakness Enumeration (CWE) codes and AWS services within the threat model. We introduce the AWS Service Coverage (ASC) metric, expressed as a percentage, to quantify the accuracy of AWS service identification. These metrics ensure LLMs capture the full threat surface, especially in complex, multi-tier deployments.

\begin{table}[htb]
\centering
\caption{CWE Coverage scores for ACSE-Eval.}
\begin{tabular}{lcccccc}
\toprule
& \multicolumn{2}{c}{\textbf{IaC}} & \multicolumn{2}{c}{\textbf{IaC + CRC}} \\
\cmidrule(lr){2-3} \cmidrule(lr){4-5}
\textbf{Model} & \textbf{0-shot} & \textbf{3-shot} & \textbf{0-shot} & \textbf{3-shot} \\
\midrule
Claude 3.5 Haiku & $15.8 \pm 1.55$ & $10.9 \pm 2.02$ & $16.9 \pm 1.76$ & $7.14 \pm 1.61$\\
Claude 3.7 Sonnet & $23.8 \pm 1.70$ & $26.9 \pm 1.84$ & $25.5 \pm 2.08$ & $27.9 \pm 2.03$\\
GPT 4o & $17.9 \pm 1.47$ & $23.6 \pm 1.99$ & $20.1 \pm 1.58$ & $25.4 \pm 1.99$\\
GPT 4.1 & $\mathbf{37.3 \pm 1.96}$ & $\mathbf{34.8 \pm 1.89}$ & $\mathbf{38.4 \pm 2.36}$ & $\mathbf{39.1 \pm 1.99}$\\
DeepSeek Chat V3 & $32.0 \pm 1.76$ & $32.0 \pm 1.76$ & $34.2 \pm 1.75$ & $31.1 \pm 1.8$\\
Gemini 2.5 Pro & $36.6 \pm 2.17$ & $32.1 \pm 2.02$ & $37.9 \pm 1.99$ & $32.0 \pm 1.88$ \\
\bottomrule
\end{tabular}
\end{table}

\begin{table}[htb]
\centering
\caption{AWS Service Coverage scores for ACSE-Eval.}
\begin{tabular}{lcccccc}
\toprule
& \multicolumn{2}{c}{\textbf{IaC}} & \multicolumn{2}{c}{\textbf{IaC + CRC}} \\
\cmidrule(lr){2-3} \cmidrule(lr){4-5}
\textbf{Model} & \textbf{0-shot} & \textbf{3-shot} & \textbf{0-shot} & \textbf{3-shot} \\
\midrule
Claude 3.5 Haiku & $23.4 \pm 1.78$ & $20.4 \pm 2.86$ & $24.3 \pm 1.72$ & $12.1 \pm 7.14$  \\
Claude 3.7 Sonnet & $34.0 \pm 2.01$ & $46.0 \pm 2.2$ & $35.7 \pm 2.06$ & $50.1 \pm 2.37$ \\
GPT 4o & $20.0 \pm 1.58$ & $37.1 \pm 2.20$ & $20.9 \pm 1.48$ & $38.0 \pm 2.11$ \\
GPT 4.1 & $44.7 \pm 2.42$ & $\mathbf{50.2 \pm 2.41}$ & $51.3 \pm 2.24$ & $\mathbf{56.0 \pm 2.57}$ \\
DeepSeek Chat V3 & $34.4 \pm 2.11$ & $45.4 \pm 2.22$ & $43.8 \pm 2.33$ & $48.2 \pm 2.44$ \\
Gemini 2.5 Pro & $\mathbf{45.6 \pm 2.08}$ & $45.8 \pm 2.24$ & $\mathbf{53.4 \pm 2.21}$ & $50.4 \pm 2.19$ \\
\bottomrule
\end{tabular}
\end{table}

Our analysis reveals several key insights into the performance of different language models. While GPT-4.1 excels in detecting CWE, its limited coverage indicates that LLMs are still in their early stages regarding security vulnerability detection. The performance dynamics shift in AWS service identification tasks, where Gemini 2.5 Pro outperforms GPT-4.1 in 0-shot scenarios, though GPT-4.1 regains its advantage when provided with additional examples. Notably, compact models such as Claude 3.5 Haiku demonstrate a trend: when presented with additional context through few-shot examples or CRC, they exhibit increased hallucination rates, leading to diminished performance metrics. This could be because compact models struggle to capture complex relationships and nuances in data due to under-fitting, while larger models are good at open domain tasks.

\section{Discussion}

Our evaluation of LLMs using ACSE-Eval reveals key insights into their strengths and limitations in cloud-native threat modeling, as well as opportunities for practical deployment and future research. We also discuss the potential for ACSE-Eval to evolve. 

\subsection{Limitations and Future Direction}



While ACSE-Eval is a promising step forward, it has limitations. First, its exclusive focus on AWS may reduce generalizability to other platforms like Azure, GCP, or hybrid environments. Future iterations should expand cross-cloud coverage. Second, the dataset captures static architecture snapshots, whereas real-world systems evolve. Future datasets should include architectural drift and lifecycle transitions.

Although the dataset includes a diverse range of vulnerabilities, it does not comprehensively cover all domain-specific or compliance-driven risks. Extending coverage to industry-specific contexts will improve realism. Furthermore, while we minimized subjectivity in evaluation through structured rubrics, future work could explore automated metrics that better reflect reasoning quality.

Several research directions emerge from our findings. Specialized fine-tuning on cloud security corpora could boost LLM performance. Multimodal approaches incorporating architecture diagrams alongside text and code could provide richer context. Interactive models that engage analysts in dialogue may yield more accurate and user-aligned outputs. Longitudinal evaluation of LLMs' ability to reason over evolving architectures is another open area. Finally, maintaining the relevance of ACSE-Eval will require regular updates to include new architectures, emerging vulnerabilities, and refined evaluation metrics.

\subsection{Ethical Considerations}

In this study, we analyze threat-modeling capabilities of LLMs on an expert-curated dataset comprising of infrastructure architecture specifications and their corresponding threat models. We strictly avoid using any personal privacy information or trade secrets that could have legal or ethical ramifications. Moreover, we ensured that all of our work complied with ethical standards and legal regulations to maintain transparency and
integrity in our research. Additionally, we understand that the nature of this work might enable certain threat actors to use our dataset as a reference for identifying and exploiting threats in real-world deployments. While that is certainly possible, we believe that this line of research is necessary to build more secure systems. Over time, research in this space will enable us to automate security and deploy more secure infrastructure without human oversight. Despite our optimistic outlook, we have taken certain steps to prevent ease of exploitation. The CDK (IaC) component of ACSE-Eval lacks the necessary meta files (such as package.json, tsconfig.json, etc)  to spin up a deployment. We have only included the files that lay out the CDK specifications and cannot be used in isolation for deployments. Moreover, the threat models in ACSE-Eval do not emphasize on implementation details and only call out high-level threats. This positions ACSE-Eval well from an ethical standpoint.

\section{Conclusion}

In this paper, we introduced ACSE-Eval, the first comprehensive dataset for evaluating LLMs' capabilities in threat modeling real-world cloud architectures. Our dataset includes 100 production-grade AWS architecture scenarios with expert-generated threat models, covering diverse application domains, complexity levels, and AWS services. Our evaluation of 6 LLMs revealed both promising capabilities and significant limitations. Our analysis demonstrates that GPT-4.1 and Gemini 2.5 Pro excel at threat identification, with Gemini 2.5 Pro performing optimally in 0-shot scenarios and GPT-4.1 showing superior results in few-shot settings. While GPT-4.1 maintains a slight overall performance advantage, Claude 3.7 Sonnet generates the most semantically sophisticated threat models but struggles with threat categorization and generalization. ACSE-Eval represents a crucial step towards automated cloud security assessment, providing security practitioners and researchers with a robust framework to enhance LLM capabilities in threat analysis and effectively safeguard modern cloud infrastructure.

%
%
%
%
\raggedright
\bibliographystyle{splncs04nat}
\bibliography{bibliography} 

\end{document}